\documentclass{ws-p8-50x6-00}
\def\lapproxeq{\lower .7ex\hbox{$\;\stackrel{\textstyle
<}{\sim}\;$}}
\def\gapproxeq{\lower .7ex\hbox{$\;\stackrel{\textstyle
>}{\sim}\;$}}
\begin{document}

\title{Status of Structure Functions and Partons}

\author{A.D. Martin}

\address{Department of Physics, University of Durham, DH1 3LE, UK\\
E-mail: A.D.Martin@durham.ac.uk}


\maketitle

\abstracts{We briefly review some of the developments in the study
of parton distributions which have occurred since DIS2000,
including discussion of uncertainties, shadowing, unintegrated and
generalized distributions.}

\section{Data and parton distribution functions}

The situation is summarised in Fig.~1, which shows the kinematic
regions in the $(x, Q^2)$ plane covered by (i) the experiments at
HERA, (ii) the fixed target deep-inelastic scattering experiments
and (iii) the single jet inclusive experiments at the Tevatron.
Roughly speaking, the fixed target experiments determine the
distributions of the $u, d, \bar{u}, \bar{d}, s, \bar{s}$ quarks
for $x \sim 0.1$ (and $Q^2 \sim 10~{\rm GeV}^2$).  The HERA $F_2$
measurements determine the sea quark distributions, and their
$\partial F_2/\partial \ln Q^2$ data determine the gluon, for $x
\sim 10^{-3}$ (and $Q^2 \sim 10~{\rm GeV}^2$).  The Tevatron jet
data determine the quarks and gluon distributions in a region
around $x \sim 0.1$ (and $Q^2 \sim 10^4~{\rm GeV}^2$).

\begin{figure}[t]
\begin{center}
%\figurebox{20pc}{15pc}{} % to have a box alone
\epsfxsize=18pc % will enlarge or reduce the postscript figures based on the xsize
\epsfbox{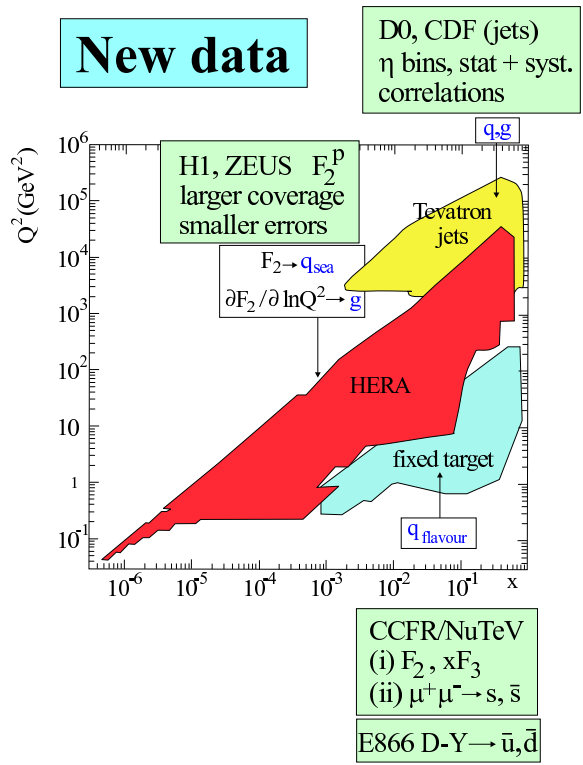} % postscript image file name
\caption{The kinematic domains probed by the various experiments,
shown together with the partons that they constrain.  The new data
that have become available since DIS2000 are also indicated.}
\end{center}
\end{figure}

In the past year new data have been available in each of the three
domains.  The H1 \cite{H1} and ZEUS \cite{ZEUS} experiments have
measured $F_2$ over a larger domain (see Fig.~1) with much
improved precision. The D0 \cite{D0} and CDF \cite{CDF}
collaborations have measured the inclusive single jet $E_T$
distribution and provided correlated statistical and systematic
errors.  In particular, the D0 experiment has measured the $E_T$
distribution in five different $\eta$ bins, out to $\eta = 3$ and
so samples partons in a much wider range of $x$ (see Fig.~1). The
NuTeV collaboration \cite{CCFR} have used $\nu, \bar{\nu}$ beams
to measure $F_2, xF_3$ with increased precision, and also, by
observing $\mu^+ \mu^-$ production, to obtain information on the
$s$ and $\bar{s}$ quark distributions; CCFR have made a
model-independent re-analysis \cite{CCFR} of their $\nu,
\bar{\nu}$ data, which removes the discrepancy for $x < 0.1$ with
NMC. The E866 collaboration \cite{E866} have observed both $pp$
and \lq\lq $pn$\rq\rq\ Drell-Yan production and further
constrained the difference of the $\bar{u}$ and $\bar{d}$
distributions.

\section{Uncertainties in parton distribution functions and observables}

The parton distributions are determined from NLO DGLAP fits to
data from about 14 diverse experiments, with typically 1500 data
points and more than 20 parameters.  There are many sources of
uncertainty.  First the statistical and systematic experimental
errors.  The latter are often not randomly distributed and,
moreover, may depend on theory.  The theoretical uncertainties
come from higher order QCD contributions, the choice of
factorization and renormalization scales, resummation corrections
(from resumming $\ln 1/x, \ln (1-x)$ terms), power law
contributions, nuclear target corrections, the choice of the
particular parametrization of the starting distributions and, in
hadronic processes, from the treatment of the underlying event.

To date, the best attempt to determine the errors on parton
distributions is due to Botje \cite{BOT}.  He includes the
statistical and systematic covariance matrices and allows for some
of the theory uncertainties.  However, the analysis is based on
five {\it massless} flavours and so is only applicable for $x >
10^{-3}$.  Moreover there is no constraint on the gluon at large
$x$ values (as would come from including Tevatron jet data in the
fit) and so the gluon is undetermined for large $x$, which feeds
down to small $x$ via the momentum sum rule.  A similar analysis,
also based on the Hessian method, has been performed by CTEQ
\cite{CTEQH}.

Recently there have been several contributions concerning the
errors on partons and observables.  Giele et al.\ \cite{GKK} have
obtained 100,000 `optimized' sets of partons, and expressed the
uncertainties as a density measure in pdf functional space.  In
principle it is `easy' to propagate errors given complete data
information, but the question of theory uncertainties has to be
addressed.  So far only $F_2^{ep}$ data have been considered;
Giele et al.\ conclude that NMC and (old) ZEUS data are
incompatible with the other $F_2^{ep}$ data sets.  This type of
analysis is in its infancy.  Clearly it is necessary to include
other types of data, e.g.\ neutrino and jet data.

It is advantageous to determine the uncertainties on observables
directly, rather than to go through the intermediate stage of
determining the parton errors.  The disadvantage is that global
analyses have to be carried out for each observable separately. In
view of the importance of $W$ and $Z$ boson hadroproduction as
luminosity monitoring processes at the Tevatron and the LHC,
attention has focused on $\sigma_{W,Z}$.  MRST(1999) \cite{MRSTW}
found an uncertainty $\Delta \sigma_W$ of $\pm 3\%$ at the
Tevatron and $\pm 5\%$ at the LHC.  In the latter case the
uncertainty is mainly due to the (conservative) error of $\pm
0.005$ assigned to $\alpha_S (M_Z^2)$.  More recently CTEQ
\cite{CTEQW} found $\Delta \sigma_W = \pm 4\%$ at the Tevatron and
$\pm 8\%$ at the LHC, taking only data errors into account.  These
surprisingly large errors are due to using older data and choosing
a large tolerance, $\chi^2 - \chi_{\rm min}^2 < 150$.  In a new
global analysis, which incorporates all the new precise data, MRST
(2001) \cite{MRST01} find $\Delta \sigma_W = \pm 2\%$ at both
Tevatron and LHC energies, when only data errors are considered.
However theory errors ($\Delta \alpha_S$ etc.) have to be
included.  Interestingly this latest analysis, which incorporates
the new Tevatron jet data \cite{D0,CDF}, gives an error of only
$\pm 20\%$ on the high $x$ gluon density.

\section{The description of $F_2$ data at low $x$}

A good description of the $F_2^{ep}$ data in the low $x$ domain
can be obtained in many different ways.  First, there are
empirical models, with very few input parameters (such as that of
Haidt \cite{HAIDT}), which give excellent descriptions down to
very low $Q^2$. Then there are dynamically-motivated
parametrizations, such as the `saturation' model of Golec-Biernat
and W\"{u}sthoff \cite{GBW} based on the $q\bar{q}$ dipole
framework.  Surprisingly, on a more comprehensive level, the NLO
DGLAP analyses \cite{H1,ZEUS,BOT,MRST01,BPZ,CTEQ} continue to give
satisfactory descriptions of the now very precise $F_2$ data, down
to remarkably low $Q^2$ and $x$.  However in these fits the gluon
tends to go valence-like or negative at low $Q^2$, $Q^2 \sim
1~{\rm GeV}^2$, indicating that the approach ceases to be valid in
this domain.  This is also reflected in the anomalous behaviour of
$F_L$ in this domain, perhaps indicating the need for $\ln (1/x)$
resummations.  In fact, a good description of $F_2$ is also
obtained in a unified approach which incorporates DGLAP and BFKL
with higher order $\ln (1/x)$ contributions \cite{KMS}. Here a
`flat' gluon is input.

The message is clear.  One low $x$ observable can be described in
many ways.  It is possible to trade effects of the different types
of perturbative evolution with different choices of the input
forms.  Of course, the situation would be changed if another
independent quantity, such as $F_L$, were precisely measured at
low $x$.

\section{BFKL and DIS}

The resummation of the $\ln (1/x)$ contributions is now known to
NLO \cite{FLCC}.  The resulting $x \rightarrow 0$ behaviour has
the form $x^{- \omega}$ with
\begin{equation}
\label{eq:a1}
 \omega \; = \; \omega_0 (1 - 6.5 \bar{\alpha}_S),
\end{equation}
where $\omega_0 = \bar{\alpha}_S 4 \ln 2$ is the LO behaviour; as
usual $\bar{\alpha}_S \equiv 3 \alpha_S/\pi$.  At first it was
thought that such large NLO corrections would mean that no stable
small $x$ predictions could be made using the BFKL procedure.
However it turns out that this is not the case.  It is possible to
identify higher-order terms and then to resum them.  Indeed
Ciafaloni et al.~\cite{CCS} carry out an {\it all-order} $\ln
(1/x)$ resummation of the following effects: (i) running
$\alpha_S$, (ii) the non-singular DGLAP terms and (iii) the
angular ordering and energy constraints.  (Thorne \cite{RST} has
made a recent detailed study of effect (i)).  The result is a
stable $x^{- \omega}$ behaviour which is consistent with
observations. In fact, prior to this, the fit of Kwiecinski et
al.~\cite{KMS}, mentioned at the end of Section~3, was based on a
unified equation which incorporates these all-order $\ln (1/x)$
contributions, where the imposition of a consistency (or
kinematic) constraint \cite{KMS2} plays a major role.

We may use a very simplified calculation to show how including
effect (iii) tames the NLO behaviour of (\ref{eq:a1}). Recall that
as we proceed along the BFKL gluon chain, we have ordering in the
longitudinal momenta, $Y > Y^\prime$, where $Y \equiv \ln (1/x)$.
To allow for the energy/angular ordering constraints, the ordering
takes the modified form
\begin{equation}
\label{eq:a2}
 Y \; > \; Y^\prime \: + \: \delta.
\end{equation}
Hence the BFKL equation for the unintegrated gluon distribution
becomes
\begin{equation}
\label{eq:a3}
 f (Y) \; = \; f_0 \: + \: \int^{Y-\delta} \: K \: f (Y^\prime) \:
 dY^\prime,
\end{equation}
or, in differential form
\begin{equation}
\label{eq:a4}
 \frac{\partial f (Y)}{\partial Y} \; = \; \omega_0 \: f (Y -
 \delta).
\end{equation}
Note that if $\delta = 0$ then $f = f_0 \exp (\omega_0 Y) \sim
x^{- \omega_0}$, as required.  In general, the solution of
(\ref{eq:a4}) is $f \sim \exp (\omega Y) \sim x^{- \omega}$, where
\begin{equation}
\label{eq:a5}
 \omega \; = \; \omega_0 \: e^{- \omega \delta} \; = \; \omega_0
 (1 - \omega_0 \delta \: + \: \ldots ).
\end{equation}
Comparison with (\ref{eq:a1}) reveals $\delta = 6.5/4 \ln 2 =
2.3$, so
\begin{equation}
\label{eq:a6}
 \omega \; = \; \omega_0 \: \exp (-2.3 \omega),
\end{equation}
which leads to a similar behaviour of $\omega$ as a function of
$\bar{\alpha}_S$ to that found in Refs.~\cite{KMS2,CCS}.  This toy
model illustrates how the inclusion of a summation of higher-order
terms stabilizes the NLO result.

A more `phenomenological' way of performing the resummation of
higher-order contributions, and achieving the perturbative
stability of the BFKL approach, has been proposed \cite{ABF}. This
incorporates the one- and two-loop terms in the anomalous
dimension, imposes momentum conservation and parametrizes the
residual ambiguity in terms of a single parameter $\omega_{\rm
eff}$ which specifies the $x^{- \omega_{\rm eff}}$ behaviour as $x
\rightarrow 0$.

\section{Parton shadowing}

Let us start with the original GLR equation \cite{GLR}, written in
terms of DGLAP evolution for the gluon density
\begin{equation}\label{eq:a7}
  \frac{\partial (xg (x, Q^2))}{\partial Y \partial \ln
  (Q^2/\Lambda^2)}\; = \; \frac{N_C \alpha_S}{\pi} xg \: - \:
  \frac{\alpha_S^2}{R^2 Q^2} \: \left [ xg \right ]^2,
\end{equation}
where, as before, $Y = \ln (1/x)$.  The first term on the
right-hand-side is the growth of $g$ due to DGLAP evolution, and
the second is the decrease due to gluon-gluon recombination.  To
gain insight into the origin of the shadowing term, note that the
number, $n$, of gluons per unit rapidity interval is $xg (x,
Q^2)$.  Moreover the gluon-gluon cross section $\hat{\sigma} (gg)
\sim \pi \alpha_S^2/Q^2$ and so
\begin{equation}\label{eq:a8}
  {\rm prob.~of~recomb.} \; \sim \; \frac{n^2 \hat{\sigma}}{\pi
  R^2} \; \sim \; \frac{\alpha_S^2}{R^2 Q^2} \: \left [xg \right
  ]^2,
\end{equation}
where $\pi R^2$ is the transverse area populated by gluons.  The
GLR equation effectively resums the `fan' diagrams generated by
the branching of QCD Pomerons.  Recently there has been much
activity in this area \cite{SHAD,KOV}, which has resulted in an
improved knowledge of the triple-Pomeron vertex.

The structure of the triple-Pomeron vertex can be extracted from
an equation \cite{KOV} for a quantity, $N (r, b, Y)$, closely
related to the cross section for the interaction of a $q\bar{q}$
dipole of transverse size $r$ with the proton target
\begin{equation}\label{eq:a9}
  \sigma (r, Y) \; = \; 2 \: \int \: d^2 b \: N (r, b, Y).
\end{equation}
$b$ is the impact parameter of the interaction.  In the
short-distance approximation $(r \ll b)$ the non-linear shadowing
equation takes the simplified form
\begin{equation}\label{eq:a10}
  \frac{\partial \tilde{N} (r, b, Y)}{\partial Y} \; = \; \frac{N_C
  \alpha_S}{\pi}\: \left \{ K \: \otimes \: \tilde{N} -
  \tilde{N}^2 \right \}
\end{equation}
in the large $N_C$ limit, where $K$ is the BFKL kernel and
\begin{equation}\label{eq:a11}
  \tilde{N} (r, b, Y) \; = \; \int_0^\infty \: \frac{d\ell}{\ell}
  \: J_0 (\ell r) \: N (\ell, b, Y).
\end{equation}

There have been two recent attempts to use (\ref{eq:a10}) to
calculate the effect of gluon shadowing \cite{KKM,LGLM}.  Both
give similar results.  Fig.~2 shows the results of Kimber et al.
We see that shadowing is small in the domain accessible to
experiments at HERA.  However almost a factor of two suppression
is anticipated for $x \sim$ few $\times 10^{-6}$.  This domain may
be accessible at the LHC for either prompt photon or high-$Q_T$
Drell-Yan production at large rapidities, both of which can
proceed via the subprocess $gq \rightarrow \gamma q$.  Another
approach \cite{BRRZ} evaluates the corrections to $\partial
F_2/\partial \ln Q^2$ due to twist-4 gluon recombination.

\begin{figure}[t]
\begin{center}
%\figurebox{20pc}{15pc}{} % to have a box alone
\epsfxsize=20pc % will enlarge or reduce the postscript figures based on the xsize
\epsfbox{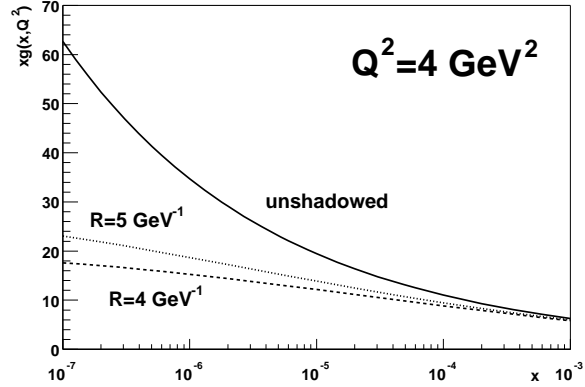} % postscript image file name
\caption{The effect of shadowing on the integrated gluon
distribution $xg (x, Q^2)$ at $Q^2 = 4~{\rm GeV}^2$, taken from
Kimber et al.~\protect\cite{KKM}.}
\end{center}
\end{figure}

If Fig.~2 shows that shadowing is small at HERA, does this rule
out the saturation models, such as \cite{GBW}?  Not necessarily;
it is hard to distinguish the $Q^2$ dependence with shadowing
present from the pure DGLAP $Q^2$ dependence, since the difference
can be removed by an adjustment of the starting distributions
\cite{CAP}. However a distinction may be possible at the LHC.  A
much better way to identify shadowing is to study the $A$
dependence of scattering on nuclei.

\section{Unintegrated parton distributions}

The natural framework with which to discuss DIS and related hard
scattering processes is to use parton distributions $f_a (x,
k_t^2, \mu^2)$, unintegrated over the transverse momentum $k_t$ of
the parton, together with the $k_t$ factorization theorem
\cite{KTFAC}.  The unintegrated distributions have the advantage
that they exactly correspond to the quantities which enter the
Feynman diagrams and therefore allow for the true kinematics of
the process even at LO.  Here, for simplicity, we will discuss the
gluon and so omit the subscript $a = g$.  The distribution $f (x,
k_t^2, \mu^2)$ depends on two hard scales --- $k_t$ and the scale
$\mu$ of the probe.  The scale $\mu$ plays a dual role.  On the
one hand it acts as the factorization scale, while on the other
hand it controls the angular ordering of the gluons emitted in the
evolution.  The {\it two-scale} distribution satisfies the CCFM
equation \cite{CCFM}, which embodies both DGLAP and BFKL
evolution.  In practice, it is complicated to solve the equation,
and, up to now, it has only proved to be practical within Monte
Carlo generators \cite{MC}.

To gain insight, recall that both DGLAP and BFKL evolution are
essentially equivalent to ordered evolution in the angles of the
emitted gluons.  In the DGLAP collinear approximation the angle
increases due to the growth of $k_t$, while in BFKL the angle
$(\theta \simeq k_t/k_\ell)$ grows due to the decreasing
longitudinal momentum fraction as we proceed along the emission
chain from the proton.  The factorization scale $\mu$ separates
the gluons associated with emission from different parts of the
process, that is from the beam and target protons (in $pp$
collisions) and from the hard subprocess.  For example, $\mu$
separates emissions from the beam (with polar angle $\theta
\lapproxeq 90^\circ$) from those of the target (with $\theta
\gapproxeq 90^\circ$), and from the intermediate gluons associated
with the hard subprocess.  This separation was proved by CCFM
\cite{CCFM} and originates from the destructive interference of
the different emission amplitudes in the angular boundary regions.
Since the evolution process is essentially controlled by one
quantity, the emission angle, we should expect to be able to
obtain the unintegrated gluon distribution $f (x, k_t^2, \mu^2)$
from a {\it single-scale} evolution equation. Indeed it is
possible to accomplish this and to follow an analytic approach
where the physical assumptions are more evident and where, in
principle, the NLO corrections can be included.  The key
observation is that the $\mu$ dependence enters only at the last
step of the evolution \cite{KMR}.

To illustrate the last-step procedure, we start from the
simplified case of pure DGLAP evolution for $G = xg$, where $g$ is
the conventional (integrated) distribution
\begin{equation}\label{eq:a12}
  \frac{\partial G (x, k_t^2)}{\partial \ln k_t^2} \; = \;
  \frac{\alpha_S}{2 \pi} \left [ \int_x^{1-\Delta} \: P_{gg} (z)
  \: G \left ( \frac{x}{z}, k_t^2 \right ) dz \: - \: G (x, k_t^2)
  \int_0^{1-\Delta} z P_{gg} (z) dz \right ].
\end{equation}
Suppose that we were to omit the virtual contribution, then the
unintegrated density would be
\begin{equation}\label{eq:a13}
  f (x, k_t^2) \; = \; \frac{\partial G (x, k_t^2)}{\partial \ln
  k_t^2} \; = \; \frac{\alpha_S}{2 \pi} \: \int_x^{1-\Delta} \:
  P_{gg} (z) \: G \left ( \frac{x}{z}, k_t^2 \right ) dz.
\end{equation}
The virtual contributions do not change the $k_t$ of the gluon and
may be resummed to give the survival probability $T$ that the
gluon remains untouched in the evolution up to the factorization
scale $\mu$
\begin{equation}\label{eq:a14}
  T (k_t, \mu) \; = \; \exp \left ( - \int_{k_t^2}^{\mu^2} \:
  \frac{dk_t^{\prime 2}}{k_t^{\prime 2}} \: \frac{\alpha_S}{2 \pi}
  \: \int_0^{1-\Delta} \: z P_{gg} (z) dz \right ),
\end{equation}
as in the Sudakov form factor.  Thus the probability to find a
gluon with transverse momentum $k_t$ (which initiates a hard
subprocess with factorization scale $\mu$) is
\begin{equation}\label{eq:a15}
  f (x, k_t^2, \mu^2) \; = \; T (k_t, \mu) \left [\frac{\alpha_S}{2
  \pi} \: \int_x^{1-\Delta} \: P_{gg} (z) \: g \left (
  \frac{x}{z}, k_t^2 \right ) \: dz \right ].
\end{equation}
It is at this last step that the unintegrated distribution becomes
dependent on $\mu$.  Angular ordering requires that the cut-off
$\Delta = k_t/(\mu + k_t)$ in (\ref{eq:a15}), and analogously in
(\ref{eq:a14}).  The last step causes the distribution to
increasingly spill over into the domain $k_t > \mu$ as $x$
decreases, as shown by the dotted curves in Fig.~3.

\begin{figure}[!htb]
\begin{center}
%\figurebox{20pc}{15pc}{} % to have a box alone
\epsfxsize=23pc % will enlarge or reduce the postscript figures based on the xsize
\epsfbox{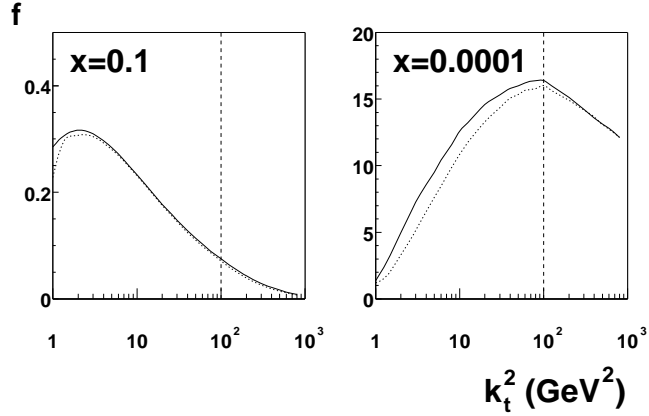} % postscript image file name
\caption{The continuous curves show the $k_t^2$ dependence of the
unintegrated gluon $f (x, k_t^2, \mu^2)$ for $x = 0.1$ and 0.0001
at $\mu = 10$~GeV obtained from a unified BFKL/DGLAP study.  The
dotted curves are obtained from DGLAP evolution.  The figure is
taken from Kimber et al.~\protect\cite{KMR}. }
\end{center}
\end{figure}

To include $\ln (1/x)$ effects, (\ref{eq:a12}) is replaced by the
unified equation of Kwiecinski et al.~\cite{KMS}, and a similar
procedure followed \cite{KMR}.  The continuous curves in Fig.~3
are obtained, which are not very different from the previous DGLAP
results.  We conclude that the main physical effects come from
angular ordering in the last step of the evolution, and not from
$\ln (1/x)$ terms.

By imposing angular ordering in both the BFKL and DGLAP terms the
integral up to $\mu^2$ of $f$ does not equal the integrated gluon.
An evaluation of $f$ which imposes the equality, but does not have
complete angular ordering, has also been made \cite{KK1}.  The
difference is a NLO effect \cite{KMR}.

\section{Generalized parton distributions}

\begin{figure}[t]
\begin{center}
%\figurebox{24pc}{15pc}{} % to have a box alone
\epsfxsize=24pc % will enlarge or reduce the postscript figures based on the xsize
\epsfbox{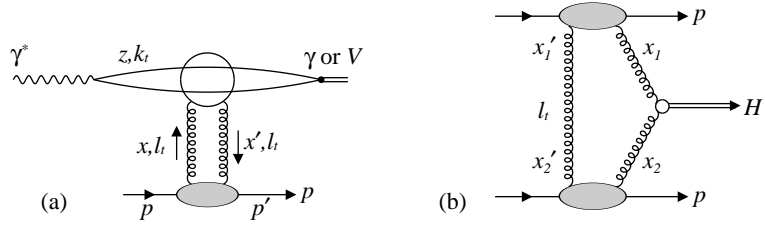} % postscript image file name
\caption{Examples of processes which depend on the generalized
(skewed) gluon distribution:  (a) DVCS and the electroproduction
of vector mesons, (b) double-diffractive Higgs production.}
\end{center}
\end{figure}

Two relevant examples, illustrating the need for generalized (or
skewed) parton distributions, are shown in Fig.~4.  The first
shows either deeply virtual Compton scattering (DVCS) or the
electroproduction of vector mesons at $t = 0$.  The second is
double-diffractive Higgs production in $pp$ collisions.  These
processes involve off-diagonal proton matrix elements $\langle
p^\prime | \ldots | p \rangle$, with longitudinal components of
momentum transfer, $x^\prime \neq x$.  We have
\begin{equation}\label{eq:a16}
  \left . \frac{d\sigma}{dt} (\gamma^* p \rightarrow \gamma p~{\rm or}~Vp)
  \right |_{t = 0} \; = \; \int dz \: \left [\int \:
  \frac{d\ell_t^2}{\ell_t^4}\: \ldots \: f (x, x^\prime; \ell_t^2,
  \mu^2) \right ]^2,
\end{equation}
where $\mu^2 = z (1 - z) Q^2 + k_t^2 + m_q^2$, and
\begin{equation}\label{eq:a17}
  \sigma (pp \rightarrow p + H + p) \; = \; \left [ \ldots \int
  \frac{d\ell_t^2}{\ell_t^4} \: f \left (x_1, x_1^\prime;
  \ell_t^2, \frac{M_H^2}{4} \right ) \: f \left (x_2, x_2^\prime;
  \ell_t^2, \frac{M_H^2}{4} \right ) \right ]^2.
\end{equation}
At small $x$ and $x^\prime << x$, applicable to these processes,
the generalized distributions are given by \cite{KH}
\begin{equation}\label{eq:a18}
  f (x, x^\prime; \ell_t^2, \mu^2) \; = \; R f(x, \ell_t^2, \mu^2) \; = \; R \:
  \frac{\partial}{\partial \ln \ell_t^2} \: \left (\sqrt{T (\ell_t,
  \mu)} \: xg (x, \ell_t^2) \right ),
\end{equation}
where the survival probability $T$ is given by (\ref{eq:a14}).
For small $x$ the ratio $R$ of the skewed to the diagonal
distribution is known \cite{SGMR}. It leads, for example, to an
enhancement of $R^2 \simeq 2$ for $\sigma (\gamma p \rightarrow
\Upsilon p)$ at HERA \cite{MRT} which seems to be required by the
data, and an enhancement of $R^4 \simeq 2$ of $\sigma (pp
\rightarrow p + H + p)$ at the LHC \cite{KH}.

Data for $\gamma^* p \rightarrow \rho_L p$ now exist from high
energies (H1,ZEUS), through intermediate energies (E665,NMC), down
to low energies (HERMES).  At high energies ($x \lapproxeq 0.01$)
the gluon mechanism shown in Fig.~4(a) dominates, whereas at low
energies $(x \gapproxeq 0.1)$ quark exchange takes over.  The
calculation of Vanderhaeghen et al.~\cite{VGG} gives a
satisfactory description.

Good data are now starting to accumulate on the classic DVCS
process, which allows generalized parton distributions to be
studied without the complications associated with the vector meson
wave function.  Allowance must be made for the Bethe-Heitler
process, but the first indications are that the theory
expectations \cite{FFS} are in good agreement with the data.

The original aim for observing DVCS was to measure the generalized
parton distributions, and then to determine the angular momentum
distributions of the partons in the proton \cite{JI}.  However
generalized distributions are interesting in their own right and
it is an active area for theoretical study.  There are many
interesting observable asymmetries.  The evolution of the
distributions is known to NLO.  In analogy to the Mellin moments
of ordinary partons, we should consider the Gegenbauer moments of
the generalized distributions.  The inverse transform is known
\cite{SHUVAEV}.  There have been attempts to calculate the
distributions in the non-perturbative region \cite{GPDNP}. An area
of present activity is the preservation of gauge invariance, which
requires a twist-3 contribution \cite{TW3}. The talk by Diehl
\cite{DIE} gives an overview of some of these developments.

\section*{Acknowledgements}

I thank Markus Diehl, Martin Kimber, Jan Kwiecinski, Misha Ryskin
and Robert Thorne for valuable help with this talk, and Rosario
Nania for arranging such a splendid Workshop in Bologna.

\end{document}